\documentclass{jetpl}
\twocolumn

\lat

\def\ntwo{${\mathcal N}=2\;$}

\def\none{${\mathcal N}=1\;$}
\def\ntwot{${\mathcal N}=(2,2)\;$}

\newcommand{\pt}{\partial}
\newcommand{\tN}{\widetilde{N}}

\newcommand{\beq}{\begin{equation}}   
\newcommand{\eeq}{\end{equation}}
\newcommand{\beqn}{\begin{eqnarray}}   
\newcommand{\eeqn}{\end{eqnarray}}

\title{Non-Abelian Vortex in
 Four Dimensions as a Critical  Superstring\,\thanks{Invited contribution, to be published in a special issue of JETP Letters.}}
 
\rtitle{Non-Abelian Vortex in
 Four Dimensions \ldots}

\sodtitle{Non-Abelian Vortex in
 Four Dimensions as a Critical  Superstring}

\author{  
M.\,Shifman$^{\,a}$ and  A.\,Yung$^{\,\,a,b,c,}$\/\thanks{e-mail: yung@thd.pnpi.spb.ru}}

\rauthor{ 
M.\,Shifman,  A.\,Yung}

\sodauthor{Shifman, Yung}

\address{
$^a${\it  William I. Fine Theoretical Physics Institute,
University of Minnesota,
Minneapolis, MN 55455, USA}\\~\\
$^{b}${\it National Research Center ``Kurchatov Institute'', 
Petersburg Nuclear Physics Institute, Gatchina, St. Petersburg
188300, Russia}\\~\\
$^{c}${\it  St. Petersburg State University,
 Universitetskaya nab., St. Petersburg 199034, Russia}}

\abstract{We discuss recent progress in describing a
 certain non-Abelian vortex string  as a critical superstring on a conifold and clarify 
 some subtle points. This particular solitonic vortex is supported in 
  four-dimensional \ntwo supersymmetric QCD with the $U(2)$ gauge group,
 $N_f=4$ quark flavors and the Fayet-Iliopoulos term.
 Under certain conditions the non-Abelian vortex can become infinitely thin and 
 can be interpreted as a critical ten-dimensional superstring. In addition to four
translational moduli the non-Abelian vortex under consideration carries six 
orientational and size moduli. 
The vortex moduli dynamics are described by a two-dimensional sigma model 
with the target space  $\mathbb{R}^4\times Y_6$
where $Y_6$ is a non-compact Calabi-Yau conifold.
The closed string states which emerge in four dimensions (4D) are identified with hadrons of 
4D bulk \ntwo QCD. It turns out that most of the states arising from the ten-dimensional graviton spectrum
are non-dynamical in 4D. A single dynamical massless hypermultiplet associated with the deformation 
of the complex structure of the conifold is found. It is interpreted as a monopole-monopole baryon 
of the 4D theory (at strong coupling).
}

\PACS{??.??.+r, ??.??.Fp}

\begin{document}

\maketitle

\section{ Introduction}

In this paper we review the recent discovery
of non-Abelian solitonic vortices \cite{HT1,ABEKY,SYmon,HT2} 
supported in certain four-dimensional \ntwo supersymmetric 
Yang-Milles theory with matter which 
behave as critical ten-dimensional superstrings.

In QCD, the Regge trajectories show almost perfect linear $J$ behavior 
($J$ stands for spin). However, in all controllable examples at weak coupling a solitonic confining string
exhibits linear behavior for the Regge trajectories only at asymptotically large spins
\cite{Yrev00,Shifman2005}.
 The reason for this is that at $J\sim 1$ the physical ``string" becomes short and thick and cannot yield
linear Regge behavior. Linear Regge trajectories at $J\sim 1$ 
have a chance to emerge only if
the string at hand
satisfies the thin-string condition \cite{SYcstring},
\beq
T\ll m^2\,,
\label{thinstring}
\eeq
where $T$ is the string tension and $m$ is a typical mass scale of the bulk fields forming the string. 
The former parameter determines the string length, while the latter determines the string thickness.
At weak coupling $g^2\ll 1$, where $g^2$ is the bulk coupling constant, we have  $m\sim g\sqrt{T}$. 
The thin-string condition (\ref{thinstring}) is therefore badly broken.

For the majority of solitonic strings in four dimensions, such as the Abrikosov-Nielsen-Olesen (ANO) vortices 
\cite{ANO}, the low-energy two-dimensional effective theory on the string world sheet -- the Nambu-Goto theory --
is not ultraviolet (UV) complete. To make the world sheet theory fully defined one has 
to take into account higher derivative corrections \cite{PolchStrom}. Higher derivative terms
run in inverse powers of $m$ and  blow up in UV making the string worldsheet ``crumpled''
\cite{Polyak86}. The blow up of higher derivative terms in
the worldsheet theory corresponds to the occurrence of a thick and short 
``string." 

The question weather one can find an example of a solitonic string which might produce  linear 
Regge trajectories at $J\sim 1$ was addressed and answered in \cite{SYcstring}. Such a string should satisfy the
thin-string condition (\ref{thinstring}). This condition means that higher derivative correction are 
parametrically small and can be ignored. If so, the low-energy world-sheet theory must be 
UV complete. This implies the  following necessary conditions:

\begin{itemize}
\item[(i)]  The low-energy world-sheet theory must be conformally invariant;

\item[(ii)] The theory must have the critical value of the Virasoro central charge.
\end{itemize}

As well-known, these conditions are satisfied by fundamental (super)string in 10D.

\vspace{2mm}

In \cite{SYcstring} it was shown  that (i) and (ii) above are met by the non-Abelian vortex string 
\cite{HT1,ABEKY,SYmon,HT2} supported in four-dimensional 
\ntwo supersymmetric QCD with the U$(N)$ gauge group, $N_f=2N$ matter hypermultiplets  and the
Fayet-Iliopoulos (FI) parameter $\xi$ \cite{FI}. The non-Abelian part of the gauge group has the
vanishing $\beta$ function. (We will need to consider $N=2$.)

The non-Abelian vortex string is 1/2
BPS saturated and, therefore,  has \ntwot supersymmetry on its worldsheet.
In addition to translational moduli characteristic of the ANO strings,
 the non-Abelian string carries orientational  moduli, as well as size moduli provided that $N_f>N$.
\cite{HT1,ABEKY,SYmon,HT2}, see \cite{Trev,Jrev,SYrev,Trev2} for reviews. Their dynamics
is described by two-dimensional sigma model with 
the target space 
\beq
\mathcal{O}(-1)^{\oplus(N_f-N)}_{\mathbb{CP}^1}\,,
\label{12}
\eeq
to which we will refer as WCP$(N,N_f-N)$ model.  For $N_f=2N$
the model becomes conformal and condition $(i)$ above is satisfied. Moreover for $N=2$ the 
dimension of orientational/size moduli space is six and they can be combined with 
four translational moduli to form a ten-dimensional space required for critical superstrings\footnote{It corresponds to $\widehat{c}=\frac{c}{3}=3$}.
Thus the second condition is also satisfied \cite{SYcstring}.
For $N=2$ the  sigma model target space is a six-dimensional
non-compact Calabi-Yau manifold $Y_6$, namely, the resolved conifold.

Given that the necessary conditions are met, a hypothesis was put forward \cite{SYcstring} 
that this non-Abelian 
vortex string does satisfy thin-string condition (\ref{thinstring}) at strong coupling regime 
in the vicinity of a critical value of $g_c^2\sim 1$. This implies  that $m(g^2) \to \infty$ at 
$ g^2\to g_c^2$.

Moreover, a version of the string-gauge duality
for the four-dimensional bulk Yang-Mills was proposed \cite{SYcstring}: at weak coupling this theory is in the Higgs phase and can be 
described in terms of (s)quarks and Higgsed gauge bosons, while at strong coupling hadrons of this theory 
can be understood as string states formed by the non-Abelian vortex string.
Later this hypothesis was further explored by studying string theory for the critical 
non-Abelian vortex \cite{2222,KSYconifold}. 
This analysis allows one to confirm and enhance the construction in \cite{SYcstring}. 

The vortices in the U$(N)$ theories under consideration
are topologically stable, therefore
the finite length strings are closed. Thus, we focus on the 
closed strings emerging from four dimensions. The goal is to  identify closed string states 
with  hadrons of the four-dimensional bulk theory. The first
step of this program, namely the identification of massless
string states was performed in \cite{2222,KSYconifold}

In particular, we identified a single matter hypermultiplet associated with the deformation 
of the complex structure of the conifold as the only 4D massless mode of the string. 
Other states arising from the ten-dimensional graviton are not dynamical
in four dimensions. In particular, 4D graviton and unwanted vector multiplets are absent.
This is due to non-compactness of the  Calabi-Yau manifold we deal with and 
non-normalizability of the corresponding modes.

It was also  discussed how the states seen in the bulk theory at weak coupling are related to
what we obtain from the string theory at strong coupling. In particular  the
 hypermultiplet associated with the deformation of the complex structure  of the conifold is interpreted 
as a monopole-monopole baryon \cite{2222,KSYconifold}.

It is worth mentioning at this point that the solitonic vortex describes only non-perturbative states. 
Perturbative states, in particular massless states associated with the Higgs 
branch of our four-dimensional theory, are present at all values of the gauge coupling and 
are not captured by the vortex string dynamics. 

\section{4D \ntwo supersymmetric QCD}

The basic bulk model we start from  is \ntwo SQCD
 with the $U(N =2)$ gauge group and $N_f=4$ massless quark hypermultiplets. 
It is described in detail in  \cite{SYmon},  see also
 the review \cite{SYrev}.
The field content is as follows. 

The \ntwo vector multiplet
consists of the $U(1)$
gauge field $A_{\mu}$ and the SU$(2)$  gauge fields $A^a_{\mu}$,
where $a=1,..., 3$, as well as their Weyl fermion superpartners plus
complex scalar fields $a$, and $a^a$ and their Weyl superpartners, respectively.

The matter sector of  the U$(N)$ theory contains
 $N_f=4$ (s)quark hypermultiplets  each consisting
of   the complex scalar fields
$q^{kA}$ and $\widetilde{q}_{Ak}$ (squarks) and
their  fermion superpartners --- all in the fundamental representation of 
the U$(2)$ gauge group.
Here $k=1,..., 2$ is the color index
while $A$ is the flavor index, $A=1,..., 4$. We also assume the quark mass parameters to vanish.

In addition, we introduce the
FI parameter $\xi$ in the U(1) factor of the gauge group.
It does not break \ntwo supersymmetry.

We will consider the bulk theory with $N_f=2N$. In this case the SU$(N)$ gauge coupling does not run
since the corresponding $\beta$ function vanishes.
Note however, that the conformal invariance of the bulk theory is explicitly broken by the FI parameter.

Let us review the vacuum structure and the excitation spectrum 
of the bulk theory assuming weak coupling, $g^2\ll 1$,
where $g^2$ is the SU$(2)$ gauge coupling.
The FI term triggers the squark condensation.
 The squark vacuum expectation values (VEV's)  are  
\beqn
\langle q^{kA}\rangle &=& \sqrt{\xi}\,
\left(
\begin{array}{cccc}
1 &  0 & 0 & 0\\
0 &  1 & 0 &  0\\
\end{array}
\right), \qquad  \langle\bar{\widetilde{q}}^{kA}\rangle= 0,
\nonumber\\[4mm]
k&=&1,2\,,\qquad A=1,...,4\, ,
\label{qvev}
\eeqn
where we present the squark fields as matrices in the color ($k$) and flavor ($A$) indices and  $\xi$ is the Fayet-Iliopoulos parameter.

The squark condensate (\ref{qvev}) results in  the spontaneous
breaking of both gauge and flavor symmetries.
A diagonal global SU$(2)$ combining the gauge SU$(2)$ and an
SU$(2)$ subgroup of the flavor SU$(4)$
group survives, however.  This is a well known phenomenon of color-flavor locking. 

Thus, the unbroken global symmetry of the bulk 
is 
\beq
  {\rm SU}(2)_{C+F}\times  {\rm SU}(2)\times {\rm U}(1)\,,
\label{c+f}
\eeq
Here SU$(2)_{C+F}$ represents a global unbroken color-flavor rotation, which involves the
first $2$ flavors, while the second SU$(2)$ factor stands for the flavor rotation of the remaining
$2$ quarks. 

The  U(1) factor in (\ref{c+f}) is the unbroken combination of the
gauge U(1) group and a U(1) subgroup of the flavor SU(4),
which rotates the first and the last two quark flavors in the opposite
directions (i.e. the charges are opposite). 
We will identify this unbroken U$(1)$ factor with a baryonic
U(1)$_{B}$ symmetry. Note, that what is usually identified as 
the baryonic U(1)$_{B}$ symmetry is a part of the U(2) gauge group 
in our bulk QCD.

Now, let us briefly discuss the perturbative excitation spectrum. 
Since both U(1) and SU($2$) gauge groups are broken by the squark condensation, all
gauge bosons become massive. In particular, the masses of the SU$(2)$ gauge bosons
are 
\beq
m\approx g\sqrt{\xi}
\label{mWc}
\eeq
at weak coupling.

As was already mentioned, \ntwo supersymmetry remains unbroken. In fact, with the
non-vanishing $\xi$, both the squarks and adjoint scalars  
combine  with the gauge bosons to form long \ntwo supermultiplets 
with eight real bosonic components \cite{VY}. 
All states appear in the representations of the unbroken global
group (\ref{c+f}), namely, in the singlet and adjoint representations
of SU$(2)_{C+F}$,
\beq
(1,\, 1,\, 0), \quad (\textbf{Adj},\, 1,\, 0),
\label{onep}
\eeq
and in the bi-fundamental representations of $\text{SU(2)}_{C+F}\times  {\rm SU}(2)$,
\beq
\left(\bar{\textbf{2}},\, \textbf{2},\, 1\right), \quad
\left(\textbf{2},\, \bar{\textbf{2}},\, - 1\right)\,.
\label{twop}
\eeq

\vspace{2mm}

The representations in (\ref{onep}) and (\ref{twop})  are labeled according to three
 factors in (\ref{c+f}). The singlet and adjoint fields are  the gauge bosons, and
 the first $2$ flavors of squarks $q^{kP}$ ($P=1,2$), together with their 
fermion superpartners. In particular, the mass of adjoint fields is given by Eq. (\ref{mWc}). 
For the detailed explanation of the U(1) charges in (\ref{onep}) and (\ref{twop}) see \cite{KSYconifold}.

The physical reason behind the fact that the (s)quarks transform in the adjoint or bi-fundamental 
representations of the global group is that their color charges are screened by the 
condensate (\ref{qvev}) and therefore they can be considered as mesons.

The bi-fundamental fields (\ref{twop}) represent the (s)quarks of the type $q^{kK}$ with $K=3,4$.
They belong to short BPS multiplets with four real bosonic components.
These fields are massless provided that the quark mass terms vanish. In fact,  in this case the 
vacuum (\ref{qvev}) in which only $N$ 
first squark flavors develop VEVs is
not an isolated vacuum. Rather, it is a root of a Higgs branch on which other flavors can also 
develop VEVs. This Higgs branch forms a cotangent bundle to the complex Grassmannian
\begin{equation}
{\cal H} = T^*\textrm{Gr}^{\mathbb{C}}_{4, 2}\,. 
\label{eq:HiggsbranchGr}
\end{equation}
whose real dimension is \cite{APS,MY}
\beq
{\rm dim}{\cal H}= 4N (N_f-N)=16.
\label{dimH}
\eeq
The above Higgs branch is non-compact and is hyper-K\"ahlerian \cite{SW2,APS}, therefore its metric cannot be 
modified by quantum corrections \cite{APS}. In particular, once the Higgs branch is present at weak coupling
we can continue it all the way into strong coupling. In principle, it can intersect with other 
branches, if present, but it cannot disappear in the theory with vanishing matter mass parameters.
We will see below that the presence of the Higgs
branch and associated massless bi-fundamental quarks has a deep impact on the non-Abelian vortex dynamics.

The BPS vortex solutions exist only on the base of the Higgs
branch which we define as a submanifold with $\tilde{q}=0$. Therefore, we will limit ourselves to the vacua which belong
to the base manifold. 


\section{World sheet model}

The presence of the global SU$(N)_{C+F}$ symmetry is the reason for
formation of non-Abelian flux tubes (vortex strings) \cite{HT1,ABEKY,SYmon,HT2}.
The most important feature of these vortices is the presence of orientational  zero modes.
In \ntwo bulk theory these strings are 1/2 BPS-saturated; hence,  their
tension  is determined  exactly by the FI parameter
\beq
T=2\pi \xi\,.
\label{ten}
\eeq

Let us briefly review the model emerging on the world sheet
of the non-Abelian string.

The translational moduli fields (they decouple from other moduli)
 in the Polyakov formulation \cite{P81} are given by the action
\beq
S_{\rm 0} = \frac{T}{2}\,\int d^2 \sigma \sqrt{h}\, 
h^{\alpha\beta}\pt_{\alpha}x^{\mu}\,\pt_{\beta}x_{\mu}
+\mbox{fermions}\,,
\label{s0}
\eeq
where $\sigma^{\alpha}$ ($\alpha=1,2$) are the world-sheet coordinates, $x^{\mu}$ ($\mu=1,...,4$) 
describe the $\mathbb{R}^4$ part  of the string
world sheet and $h={\rm det}\,(h_{\alpha\beta})$, where $h_{\alpha\beta}$ is the world-sheet metric 
which is understood as an
independent variable.

If $N_f=N$  the dynamics of the orientational zero modes of the non-Abelian vortex, which become orientational moduli fields 
 on the world sheet, is described by two-dimensional
\ntwot-supersymmetric $CP(N-1)$ model.

If one adds extra quark flavors, non-Abelian vortices become semilocal.
They acquire size moduli \cite{AchVas}.  
In particular, for the non-Abelian semilocal vortex   in 
addition to  orientational zero modes of the vortex
string $n^P$ ($P=1,2$), there are   size moduli   
$\rho^K$ ($K=1,2$) \cite{AchVas,HT1,HT2,SYsem,Jsem,SVY}.  

The  gauged formulation of the effective world sheet theory for orientational and size moduli is as follows \cite{W93}. One introduces
 the $U(1)$ charges $\pm 1$, namely $+1$ for $n$'s and $-1$ for $\rho$'s, 
\beqn
S_{\rm 1} &=& \int d^2 \sigma \sqrt{h} \left\{ h^{\alpha\beta}\left(
 \tilde{\nabla}_{\alpha}\bar{n}_P\,\nabla_{\beta} \,n^{P} 
 +\nabla_{\alpha}\bar{\rho}_K\,\tilde{\nabla}_{\beta} \,\rho^K\right)
 \right.
\nonumber\\[3mm]
&+&\left.
 \frac{e^2}{2} \left(|n^{P}|^2-|\rho^K|^2 -\beta\right)^2
\right\}+\mbox{fermions}\,,
\label{wcp}
\eeqn
where 
\beq 
\nabla_{\alpha}=\pt_{\alpha}-iA_{\alpha}\,, \qquad \tilde{\nabla}_{\alpha}=\pt_{\alpha}+iA_{\alpha}
\eeq
 and $A_\alpha$ is an auxiliary gauge field.
 The limit $e^2\to\infty$ is implied. Equation (\ref{wcp}) represents the   $WCP(2,2) $ model.\footnote{Both the orientational and the size moduli
have logarithmically divergent norms, see e.g.  \cite{SYsem}. After an appropriate infrared 
regularization, logarithmically divergent norms  can be absorbed into the definition of 
relevant two-dimensional fields  \cite{SYsem}.
In fact, the world-sheet theory on the semilocal non-Abelian string is 
not exactly the $WCP(N,\tN)$ model \cite{SVY}, there are minor differences. The actual theory is called the $zn$ model. Nevertheless it has the same infrared physics as the GLSM (\ref{wcp}) \cite{KSVY}.} 

The total number of real bosonic degrees of freedom in (\ref{wcp}) is six, 
where we take into account the constraint
imposed by $D$-term. Moreover, one U(1) phase is gauged away. These six internal degrees of freedom 
are combined with four translational moduli from (\ref{s0}) to form a ten-dimensional space needed 
for a superstring to be critical.

In the semiclassical approximation the coupling constant $\beta$ in (\ref{wcp}) is related to 
the bulk $SU(2)$ gauge coupling 
$g^2$ via \cite{SYrev}
\beq
\beta\approx \frac{4\pi}{g^2}\,.
\label{betag}
\eeq
Note that the first (and the only) coefficient is the same for the bulk and 
world-sheet beta 
functions and vanishes. This ensures that our world sheet theory is conformal
invariant.

The total world-sheet action is
\beq
S=S_0+S_1\,.
\label{stringaction}
\eeq

The global symmetry of the world-sheet sigma model (\ref{wcp}) 
is
\beq
 {\rm SU}(2)\times {\rm SU}(2)\times {\rm U}(1);
\label{globgroup}
\eeq
it is the same as the unbroken global group of the bulk theory (\ref{c+f}). 
The fields $n$ and $\rho$ 
transform in the following representations:
\beq
n:\quad (\textbf{2},\,0,\, 0), \qquad \rho:\quad (0,\,\textbf{2},\, 1)\,.
\label{repsnrho}
\eeq


\section{Thin string regime}

As we know \cite{ArgPlessShapiro,APS}  the bulk theory at hand possesses a strong-weak coupling 
duality\,\footnote{Argyres et al. proved this duality for $\xi =0$. It should allow one to 
study the bulk theory at strong coupling in terms of weakly coupled dual theory  
 at $\xi\neq 0$ too.}
\beq
\tau\to \tau_{D} = -\frac{1}{\tau}\,,\qquad \tau = i\frac{4\pi}{g^2} +\frac{\theta_{4D}}{2\pi}\,,
\label{bulkduality}
\eeq 
 where $\theta_{4D}$ is the 4D $\theta$-angle.

The world-sheet model (\ref{wcp}) is also obviously
self-dual under the reflection of the coupling constant $\beta$,
\beq
\beta \to  \beta_D = -\beta\,.
\label{CPduality}
\eeq
Under this duality the orientational and size moduli $n^P$ and $\rho^K$ interchange.

The 2D coupling constant $\beta$ can be naturally complexified if we include the $\theta$ term in the 
action of the $\mathbb{CP}^{N-1}$ model, 
$$\beta \to \beta + i\,\frac{\theta_{2D}}{2\pi}\,.$$ 
Given the complexification of $\beta$ we have a generalization of (\ref{CPduality}) to complex values of the coupling. 

The exact relation between 4D and 2D couplings reads
\beq
\exp{(-2\pi\beta)} = - h(\tau)[h(\tau) +2],
\label{taubeta}
\eeq
where the function $h(\tau)$ is a special modular function of $\tau$ defined in terms of $\theta$-functions: $h(\tau)=\theta_1^4/(\theta_2^4-\theta_1^4)$.
 This function enters the Seiberg-Witten curve for our 4D QCD
 \cite{ArgPlessShapiro,APS}. The derivation of the relation (\ref{taubeta}) will be presented elsewhere \cite{Komarg,tobepub}.

Note, that the 4D self-dual point $g^2=4\pi$ is mapped onto the 2D self-dual point $\beta=0$.

It was conjectured in \cite{SYcstring} that thin-string condition (\ref{thinstring}) is in fact satisfied
in this theory at strong coupling limit $g^2_c\sim 1$. 
The conjecture is equivalent
to the assumption that the mass of quarks and gauge bosons $m$ 
has a singularity  as a function of $g^2$. If we assume for simplicity that there is only one singular point, then by
symmetry, a 
natural choice is the self-dual point $\tau_c =i$ or $g^2_c =4\pi$.  This gives
\beq
m^2  \to \xi\times 
\left\{
\begin{array}{ccc}
g^2, & g^2\ll 1& \\
\infty, & g^2\to 4\pi& \\
16\pi^2/g^2,& g^2\gg 1 & \\
\end{array},
\right.
\label{msing}
\eeq
where the dependence of $m$  at small and large $g^2$ follows from the tree-level formula (\ref{mWc}) and
duality (\ref{bulkduality}).

Thus we expect that the  singularity of mass $m$ lies at $\beta=0$. This  is the point where the non-Abelian string 
becomes infinitely thin,  higher derivative terms can be neglected and the theory of the non-Abelian 
string reduces to (\ref{stringaction}). The point $\beta=0$ is a natural choice because at this point
 we have a regime change in the 2D sigma model  {\em per se}. 
This is the point where the resolved conifold defined by the $D$-term in
(\ref{wcp}) develops a conical singularity \cite{NVafa}.

The term ``thin string'' should be understood with care. 
As was mentioned previously,  the target space of our sigma model is  {\em non-compact}, see  (\ref{wcp}). 
Since the non-compact
string moduli $\rho^K$ have the string-size interpretation one might think that at large $|\rho |$ our
 string is not thin. Note, that by the thin-string condition 
(\ref{thinstring}) we mean  that the string core  is thin, and higher-derivative corrections which run 
in powers of  ${\pt^2}/{m^2}$  are negligible.

Note that  there are massless states in the bulk theory namely bifundamental quarks (\ref{twop}) 
which give rise to the continuous spectrum.
Most of  these light modes are {\em not} localized on the string and do not participate in the string dynamics. 
The only zero modes which are localized
 (in addition to the  translational modes) are the size and the orientational modes \cite{SYsem} indicated in
(\ref{wcp}). They have a logarithmically divergent norm while all other light modes are power-like non-normalizable in 
the infrared.  All other localized modes are massive, 
with mass $\sim m$. Integrating out these massive modes leads
to higher-derivative corrections running in powers of  ${\pt^2}/{m^2}$. They are negligible if   
$m$ is large, see (\ref{thinstring}). We do {\em not} integrate out zero modes.


\section{4D bulk supersymmetry from the critical non-Abelian vortex string}

The critical string 
discovered in \cite{SYcstring} must lead to
${\mathcal N}=2$ supersymmetric spectrum in our bulk four-dimensional QCD. 
This is {\em a priori} clear
because  our starting basic theory is ${\mathcal N}=2$. The question  is how this symmetry emerges 
from the string world-sheet model. 

Non-Abelian  vortex string  is 1/2 BPS; therefore, out of eight bulk supercharges it preserves four.
These four supercharges form ${\mathcal N} =(2,2)$  on the world sheet which 
is necessary to have \ntwo space-time supersymmetry \cite{Gepner,BDFM}. 

Note that this conclusion is  based on the fact that we have only
closed strings in our theory. If open strings were present they would break 4D supersymmetry down to \none\!\!\!, 
which would contradict the  \ntwo supersymmetry initially present in our bulk QCD. 

Another question to ask is  whether the string theory (\ref{stringaction}) 
belongs to Type IIA or IIB? We started with \ntwo supersymmetric QCD
which is a vector-like theory and preserves 4D parity.  Therefore we expect that 
the closed string spectrum
in this theory should respect 4D parity. 

On the other hand we know that Type IIB string
is a chiral theory and breaks parity while Type IIA string theory 
is left-right symmetric and conserves parity \cite{GSW}. Thus we expect that the string
theory of non-Abelian vortex is of Type IIA. This expectation is confirmed by 
studying how parity transformation acts on the 2D fermions \cite{KSYconifold}.

\section{Spectrum: 4D graviton and vector fields}

In this section we review the 4D reduction of our string theory
\cite{2222,KSYconifold}.

As was mentioned above, our target space
is $\mathbb{R}^4\times WCP(2,2)= \mathbb{R}^4\times Y_6$ where $Y_6$ is a non-compact  
Calabi-Yau conifold, which in fact, at $\beta\neq 0$  is a resolved conifold, 
see \cite{NVafa} for a review. 
We will consider the string theory (\ref{stringaction}) at small $\beta$.

Strictly speaking, at small $\beta$ the sigma model quantum corrections blow up. In other words,
 we can say that at small $\beta$ the gravity approximation does not work. However, 
 for the massless states, we can do the computations at large $\beta$ where supergravity approximation is valid and 
 then extrapolate to strong coupling. This is the reason why now we will limit ourselves only to massless states.
In the sigma model language the latter corresponds to chiral primary operators. They are protected by
\ntwot world sheet supersymmetry and their masses are not lifted by quantum corrections. However,
kinetic terms (the K\"ahler potentials) can be corrected. 

In this  section we show that all massless 4D string modes at $\beta\neq 0$ have infinite norm over the conifold and thus  decouple. In particular, the 4D-graviton and 4D vector fields
are absent.

The massless 10D boson fields of type IIA string theory are graviton, dilaton and 
antisymmetric tensor $B_{MN}$ in the NS-NS sector. In the R-R sector type IIA string gives rise to
 one-form and three-form.
(Above $M,N=1,...,10$ are 10D indices). We start from massless 10D graviton and study what states it can
produce in four dimensions. In fact, 4D states coming from other massless 10D fields listed above can be recovered
from \ntwo supersymmetry in the bulk, see for example \cite{Louis}. We will follow the standard 
string theory approach well developed for compact Calabi-Yau spaces. The only novel aspect
of the case at hand is that for each 4D state we have to check whether its wave function is normalizable on 
$Y_6$ keeping in mind that this space is non-compact.

Massless 10D graviton is represented by fluctuations of the metric  $\delta G_{MN} = G_{MN} - G_{MN}^{(0)}$,
where $G_{MN}^{(0)}$ is the metric on $\mathbb{R}^4\times Y_6$ which has a block form since $R^4$ and 
$Y_6$ are factorized. Moreover, 
$\delta G_{MN}$ should satisfy the Lichnerowicz equation
\beq
D_A D^A \delta G_{MN} + 2R_{MANB}\delta G^{AB}=0\,.
\label{10DLich}
\eeq
Here $D^A$ and $R_{MANB}$ are the covariant derivative and the Riemann tensor corresponding to 
  the background
metric $ G_{MN}^{(0)}$, where the gauge condition $D_A\delta G_{N}^A -\frac12 D_N\delta G_{A}^A=0$ is imposed. Given the block form
of  $ G_{MN}^{(0)}$,  only the six-dimensional part $R_{ijkl}$ of $R_{MANB}$ is nonzero while the
operator $ D_A D^A$ is given by $ D_A D^A= \pt_{\mu}\pt^{\mu} + D_i D^i$, where 
the indices $\mu,\nu =1,...,4$ 
and $i,j=1,...,6$
belong to flat 4D space and $Y_6$, respectively, and we use 4D metric with the 
diagonal entries $(-1,\,1,\,1,\,1)$.

Next, we search  for the solutions of (\ref{10DLich})
subject to  the block form of $\delta G_{MN}$,
\beqn
&&\delta G_{\mu\nu}=\delta g_{\mu\nu}(x)\,\phi_6(y), \quad \delta G_{\mu i}=B_{\mu}(x)\,V_i(y), 
\nonumber\\
&&\delta G_{ij}=\phi_4(x)\,\delta g_{ij}(y)\,,
\label{factor}
\eeqn
where $x_{\mu}$ and $y_i$ are the coordinates on $R^4$ and $Y_6$, respectively. We see that
$\delta g_{\mu\nu}(x)$, $B_{\mu}(x)$
and $\phi_4(x)$ are the 4D graviton,  vector  and  scalar fields, while $\phi_6(y)$, $V_i(y)$ and 
$\delta g_{ij}(y)$
are the fields on $Y_6$. 

\vspace{2mm}

Tor the fields $\delta g_{\mu\nu}(x)$, $B_{\mu}(x)$
and $\phi_4(x)$ to be dynamical in 4D, the fields $\phi_6(y)$, $V_i(y)$ and $\delta g_{ij}(y)$ 
should have finite norm
 over the six-dimensional internal space $Y_6$. Otherwise 4D fields will appear with infinite
kinetic term coefficients, and, hence, will decouple.

Symbolically the Lichnerowicz equation (\ref{10DLich}) can be written as
\beq
(\pt_{\mu}\pt^{\mu} + \Delta_6)\,g_4(x)g_6(y)=0,
\label{symbol}
\eeq
where $\Delta_6$ is the two-derivative operator from (\ref{10DLich}) reduced to $Y_6$, while
$g_4(x)g_6(y)$ symbolically denotes the factorized form in (\ref{factor}). If we expand
$g_6$ in eighenfunctions 
\beq
-\Delta_6 g_6(y)=\lambda_6 g_6(y)
\label{lambda}
\eeq
the  eighenvalues  $\lambda_6$ will play the role of the mass squared of the 4D states. 
Here  we will be interested only in the $\lambda_6=0$ eigenfunctions. Solutions of the
equation 
\beq
\Delta_6 g_6(y)=0
\label{g6equation}
\eeq
 for the Calabi-Yau manifolds
are given by elements of Dolbeault cohomology group $H^{(p,q)}(Y_6)$, where 
$(p,q)$ denotes the numbers of holomorphic and anti-holomorphic indices of the form. The numbers 
of these forms 
$h^{(p,q)}$ are called the Hodge numbers for a given $Y_6$. 

 Due to the fact that $h^{(0,0)}=1$, 
we can easily find (the only!) solution for the 4D graviton:  it is a constant on $Y_6$. This 
mode is non-normalizable. Hence,  no 4D graviton
emerges from our string.

This is a good news. We started from \ntwo QCD in 
four dimensions without gravity and, therefore, do not expect 4D graviton to appear as a closed 
string state.\footnote{The alternative option that the
massless 4D spin-2 state  has no interpretation in terms of 4D gravity is ruled 
out by Weinberg-Witten theorem \cite{ww}.}

The infinite norm of the graviton wave function on $Y_6$  rules out  
other  4D states of the \ntwo 
gravitational and tensor multiplets: the vector field, the dilaton,  the antisymmetric tensor and two scalars
coming from the 10D three-form.

Now, let us pass to the components of the 10D graviton $\delta G_{\mu i}$ which give rise 
to a vector field in 4D. The very possibility of having vector fields is due to continuous symmetries
of $Y_6$.  Namely for our $Y_6$  we expect seven Killing vectors associated with symmetry (\ref{globgroup}), which  obey the equations
\beq
D_iV^m_j +D_jV^m_i =0, \qquad m=1,...,7\,.
\label{Killingeq}
\eeq
For Calabi-Yau this implies the equation $D_jD^j V_i^m=0$, which
 for compact manifolds is equivalent to the statement 
that $V_i$ is a covariantly constant vector, $D_jV_i=0$. This is impossible for  Calabi-Yau
 manifolds with the SU(3) holonomy \cite{GSW}. 
 
 For non-compact $Y_6$ we expect the presence of seven Killing vectors associated with symmetry (\ref{globgroup}). However it is easy to see that the $V_i^m$ fields produced by 
rotations of the $y_i$ coordinates 
do not fall-off at large values of $y_i^2$ (where the metric tends to flat). Thus, they 
are non-normalizable, and the associated  4D vector fields $B_{\mu}(x)$ are absent. 


\section{Deformations of the conifold metric.}

It might seem that our 4D string does not produce 
massless 4D states at all. 
This is a wrong impression. At the selfdual value of $\beta=0$ it does! 

Consider the last option in (\ref{factor}), scalar 4D fields. 
In this case the appropriate  Lichnerowicz equation on $Y_6$ reduces
to   
\beq
D_k D^k \delta g_{ij} + 2R_{ikjl}\delta g^{kl}=0\,.
\label{6DLich}
\eeq
Solutions of this equation for the Calabi-Yau spaces reduce to  deformations of the K\"ahler form or deformations
of complex structure \cite{NVafa,GukVafaWitt}. For a generic Calabi-Yau manifold the numbers of these deformations are given by 
$h^{(1,1)}$ and $h^{(1,2)}$, respectively. Before describing these deformations we will briefly
review  conifold geometry.

The target space of the sigma model (\ref{wcp}) is defined by the $D$ term condition
\beq
|n^P|^2-|\rho^K|^2 = \beta.
\label{Fterm}
\eeq 
Also, a U(1) phase can be  gauged away. 
We can construct U(1) gauge-invariant ``mesonic'' variables
\beq
w^{PK}= n^P \rho^K.
\label{w}
\eeq
In terms of these variables the condition (\ref{Fterm}) can be written as
${\rm det}\, w^{PK} =0$, or
\beq
\sum_{\alpha =1}^{4} w_{\alpha}^2 =0,
\label{coni}
\eeq
where $w^{PK}=\sigma_{\alpha}^{PK}w_{\alpha}$, and $\sigma$-matrices are  $(1,-i\tau^a)$, $a=1,2,3$.
Equation (\ref{coni}) define the conifold -- a cone with the section $S_2\times S_3$.
It has the K\"ahler Ricci-flat metric and represents a non-compact Calabi-Yau manifold \cite{Candel,W93,NVafa}. 

At $\beta =0$ the conifold develops a conical singularity, so both $S_2$ and $S_3$  can shrink to zero.
The conifold singularity can be smoothed in two different ways: by a deformation 
of the K\"ahler form or by a deformation of the 
complex structure. The first option is called the resolved conifold and amounts to introducing 
a non-zero $\beta$ in (\ref{Fterm}). This resolution preserves 
the K\"ahler structure and Ricci-flatness of the metric. 
If we put $\rho^K=0$ in (\ref{wcp}) we get the $CP(1)$ model with the $S_2$ target space
(with the radius $\sqrt{\beta}$). The explicit metric for the resolved conifold can be found in 
\cite{Candel,Zayas,Klebanov}. The resolved conifold has no normalizable zero modes. In particular, 
it is demonstrated in \cite{KSYconifold} that the 4D scalar $\beta$ associated with 
deformation of the K\"ahler form is not normalizable.

As  explained in \cite{GukVafaWitt,KSYconifold}, non-normalizable 4D modes can be interpreted as (frozen) 
coupling constants in the 4D bulk theory. 
The $\beta$ field is the most straightforward example of this, since the 2D coupling $\beta$ is
 related to the 4D coupling, see Eq. (\ref{taubeta}).

If $\beta=0$ another option exists, namely a deformation 
of the complex structure \cite{NVafa}. 
It   preserves the
K\"ahler  structure and Ricci-flatness  of the conifold and is 
usually referred to as the deformed conifold. 
It  is defined by deformation of Eq.~(\ref{coni}), namely   
\beq
\sum_{\alpha =1}^{4} w_{\alpha}^2 = b\,,
\label{deformedconi}
\eeq
where $b$ is a complex number.
Now  the $S_3$ can not shrink to zero, its minimal size is 
determined by
$b$. The explicit metric on the deformed conifold is written down in \cite{Candel,Ohta,KlebStrass}.
 The parameter $b$ becomes a 4D complex scalar field.
The effective action for  this field is
\beq
S(b) = T\int d^4x \,h_{b}|\pt_{\mu} b|^2,
\label{Sb}
\eeq
where the metric $h_{b}(b)$ is given by the normalization integral over the conifold $Y_6$,
\beq
h_{b} = \int d^6 y \sqrt{g} g^{li}\left(\frac{\pt}{\pt b} g_{ij}\right)
g^{jk}\left(\frac{\pt}{\pt \bar{b}} g_{kl}\right),
\label{hbgen}
\eeq
and $g_{ij}(b)$ is the deformed conifold metric.

The metric $h_b$ is calculated  in \cite{KSYconifold} by two different methods, one of them \cite{GukVafaWitt} was 
kindly pointed out to us by Cumrun Vafa. Another one uses the explicit metric on the deformed conifold \cite{Candel,Ohta,KlebStrass}. The norm of
the $b$ modulus turns out to be  logarithmically divergent in the infrared.
The modes with logarithmically divergent norm are on the borderline between normalizable 
and non-normalizable modes. Usually
such states are considered as ``localized'' on the string. We follow this rule.  We can
 relate this logarithmic behavior with the marginal stability of the $b$ state, see \cite{KSYconifold}.
This scalar mode is localized on the string in the same sense as the orientational 
and size zero modes 
are localized on the vortex-string  solution.
   
   In type IIA superstring the complex scalar associated with deformations of the complex structure of 
   the Calabi-Yau
space enters
as a 4D hypermultiplet. Thus our 4D scalar $b$ is a part of a hypermultiplet. Another complex scalar
$\tilde{b}$ comes from 10D three-form, see \cite{Louis} for a review. Together they form the bosonic
content
 of a 4D \ntwo hypermultiplet. The fields $b$ and $\tilde{b}$ being massless can develop VEVs. Thus, 
we have a new Higgs branch in the bulk which is developed only at the self-dual value of 
coupling constant $g^2=4\pi$. The bosonic part of the full effective action for the
 $b$ hypermultiplet takes the following form \cite{KSYconifold}:
\beq
S(b) = T\int d^4x \left\{|\pt_{\mu} b|^2 +|\pt_{\mu} \widetilde{b}|^2 \right\}\,
\log{\frac{T^4 L^8}{|b|^2+| \widetilde{b}|^2}}\,,
\label{Sbtb}
\eeq
where $L$ is the  size of $R^4$ introduced as an infrared regularization of logarithmically divergent
norm of $b$-field.

The logarithmic metric in (\ref{Sbtb}) in principle can receive both perturbative and 
non-perturbative quantum corrections. However, for \ntwo  theory the non-renormalization
theorem of \cite{APS} forbids the  dependence of the Higgs branch metric  on the 4D coupling 
constant $g^2$.
Since the 2D coupling $\beta$ is related to $g^2$ we expect that the logarithmic metric in (\ref{Sbtb})
will stay intact.

The presence of the 
``non-perturbatively emergent'' Higgs branch
 at the self-dual point  $g^2=4\pi$ at strong coupling is a  successful test of our picture. 
A hypermultiplet is a BPS state. If it were present in a continuous region of $\tau$ 
at strong coupling it could be continued all the way to the weak coupling domain where its presence would 
contradict the  quasiclassical analysis of bulk QCD, see \cite{SYrev,KSYconifold}.


\section{ String states interpretation in the bulk.}

To find the place for the massless scalar hypermultiplet
we obtained as the only massless string excitation at the critical point $\beta=0$ let us first 
examine the weak coupling domain $g^2\ll 1$.

Since squarks develop condensates (\ref{qvev}),
the non-Abelian vortices confine monopoles, see \cite{SYrev} for details.
The elementary  monopoles are junctions of two distinct elementary 
non-Abelian strings \cite{T,SYmon,HT2}. As a result
in the bulk theory  we have 
monopole-antimonopole mesons in which a monopole and a antimonopole are 
connected by two confining strings.
 For the U(2) gauge group we have also  ``baryons" consisting
of two monopoles connected by two confining strings, see \cite{SYrev} for further details.

The monopoles acquire quantum numbers with respect to the global group (\ref{c+f})
of the bulk theory. Indeed, in the world-sheet
model on the vortex-string, the  confined monopole are seen as  
kinks interpolating between two different vacua \cite{T,SYmon,HT2}. These  kinks  are 
described at strong coupling by the $n^P$ and $\rho^K$ fields \cite{W79,HoVa}
(for $WCP(N,\tN)$ models see \cite{SYtorkink}) and therefore transform in 
the fundamental representations of global group (\ref{globgroup}) of the world sheet theory, 
which coincides with (\ref{c+f}). 

As a result, the monopole-antimonopole mesons and baryons can be  either singlets or triplets
of both $SU(2)$ global groups, as well as in the bi-fundamental representations. With respect
to the baryonic $U(1)_B$ symmetry which we define as a U(1) factor in the 
global group (\ref{c+f}), the mesons have charges $Q_{B}({\rm meson})=0,1$ while the
baryons can have charges
\beq
Q_{B}({\rm baryon})=0,1,2\,.
\label{Bbaryons}
\eeq
 All the above non-perturbative stringy states are heavy, with mass of the
order of $\sqrt{\xi}$, and can decay into screened quarks, which are lighter, and, eventually, into
massless bi-fundamental screened quarks.

Now we return from weak to strong coupling and go to the  self-dual point $\beta=0$. 
At this point a new `exotic' Higgs branch opens up which 
is parameterized by the massless hypermultiplet -- the $b$ state,  associated with the deformation of the 
complex structure of the conifold. It can be interpreted as a baryon
constructed of two monopoles connected by two strings.  To see this we note that the complex 
parameter $b$ (promoted to a 4D scalar field) is singlet with respect to two $SU(2)$ factors 
of the global world-sheet group (\ref{globgroup})
while its baryonic charge is $Q_{B}=2$ \cite{KSYconifold}.

Since the world sheet and the bulk global groups
are identical we can conclude that our massless $b$ hypermultiplet is a stringy monopole-monopole baryon.

Being massless it is marginally stable at $\beta=0$ and can decay into pair of massless bi-fundamental
quarks in the singlet channel with the same baryon charge $Q_{B}=2$. The 
$b$ hypermultiplet does not exist at non-zero ${\beta}$.

\vspace{2mm}

 {\em Acknowledgments.}---We are very grateful to Igor Klebanov and Cumrum Vafa for very useful correspondence and insights, and 
to Nathan Berkovits, Alexander Gorsky, David Gross, Igor Klebanov, Zohar Komargodski, Peter Koroteev, Andrei Mikhailov and 
Shimon Yankielowicz for helpful comments.

The work of M.S. is supported in part by DOE grant DE-SC0011842. 
The work of A.Y. was  supported by William I. Fine Theoretical Physics Institute,   
University of Minnesota,
by Russian Foundation for Basic Research Grant No. 13-02-00042a and by Russian State Grant for
Scientific Schools RSGSS-657512010.2. The work of A.Y. was supported by the Russian Scientific Foundation 
Grant No. 14-22-00281.

\end{document}